\documentclass[a4paper,fleqn,usenatbib]{mnras}

\usepackage[T1]{fontenc}
\usepackage{ae,aecompl}

\usepackage{graphicx}	
\usepackage{amsmath}	
\usepackage{amssymb}	

\title[No superluminal motion in Sw~J1644$+$57]{No apparent superluminal motion in the first-known jetted tidal disruption event {\it{Swift}}~J1644$+$5734}

\author[J. Yang et al.]{J.~Yang,$^{1,2,3}$\thanks{E-mail: jun.yang@chalmers.se}
Z.~Paragi,$^{3}$
A.J.~van~der~Horst,$^{4}$ 
L.I.~Gurvits,$^{3,5}$
R.M.~Campbell,$^{3}$ 
\newauthor
D.~Giannios,$^{6}$  
T.~An$^{2,7}$ 
and S.~Komossa$^{8}$
\\
$^{1}$Department of Earth and Space Sciences, Chalmers University of Technology, Onsala Space Observatory, 439\,92 Onsala, Sweden \\
$^{2}$Shanghai Astronomical Observatory, Chinese Academy of Sciences, 200030 Shanghai, P.R. China \\
$^{3}$Joint Institute for VLBI ERIC (JIVE), Postbus~2, 7990\,AA Dwingeloo, The Netherlands \\
$^{4}$Department of Physics, The George Washington University, 725 21st Street NW, Washington, DC 20052, USA \\
$^{5}$Department of Astrodynamics \& Space Missions, Delft University of Technology, 2629 HS Delft, Delft, The Netherlands \\
$^{6}$Department of Physics and Astronomy, Purdue University, 525 Northwestern Avenue, West Lafayette, IN 47907, USA \\
$^{7}$Key Laboratory of Radio Astronomy, Chinese Academy of Sciences, 210008 Nanjing, P.R. China \\
$^{8}$Max-Planck-Institut f\"ur Radioastronomie, Auf dem H\"ugel 69, 53121 Bonn, Germany 
}

\date{Accepted 2016 Mmm. Received 2016 Mmm. ; in original form 2016 Mmm. dd}

\pubyear{2016}

\begin{document}
\label{firstpage}
\pagerange{\pageref{firstpage}--\pageref{lastpage}}
\maketitle

\begin{abstract}
The first-known tidal disruption event (TDE) with strong evidence for a relativistic jet -- based on extensive multi-wavelength campaigns -- is {\it{Swift}}~J1644$+$5734. In order to directly measure the apparent speed of the radio jet, we performed very long baseline interferometry (VLBI) observations with the European VLBI network (EVN) at 5~GHz. Our observing strategy was to identify a very nearby and compact radio source with the real-time e-EVN, and then utilise this source as a stationary astrometry reference point in the later five deep EVN observations. With respect to the in-beam source FIRST~J1644$+$5736, we have achieved a statistical astrometric precision about 12~$\umu$as (68\,\% confidence level) per epoch. This is one of the best phase-referencing measurements available to date. No proper motion has been detected in the {\it{Swift}}~J1644$+$5734 radio ejecta. We conclude that the apparent average ejection speed between 2012.2 and 2015.2 was less than 0.3$c$ with a confidence level of 99\,\%. This tight limit is direct observational evidence for either a very small viewing angle or a strong jet deceleration due to interactions with a dense circum-nuclear medium, in agreement with some recent theoretical studies.  
\end{abstract}

\begin{keywords}
galaxies: jets -- galaxies: individual: Swift~J1644$+$5734 -- radio continuum: galaxies.
\end{keywords}


\section{Introduction}
\label{sec1}

The source {\it{Swift}}~J1644$+$5734 is the first-known transient that has extremely luminous non-thermal emission from radio to $\gamma$-ray wavelengths \citep[][]{bur11, blo11, zau11} after a star was disrupted by the gravity of a supermassive black hole \citep[see a recent review by][]{kom15}. The tidal disruption event (TDE) took place in a star-forming galaxy at $z=0.354$ \citep{lev11}. The evolution of the X-ray light curve up to 500 days roughly followed a $t^{-5/3}$ power law decline as expected for the fallback rate of tidally disrupted material \citep[e.g.][]{lod09}. There was also fast variability on time scales down to $\sim$100 seconds \citep{blo11} in X-rays. However, the radio emission had no rapid variability and evolved in a different way. At centimetre wavelengths, it brightened steadily in the first 10 days, then more slowly over the next 100 days \citep{ber12}. Most likely, the radio emission was produced by a region different from the X-ray emission site, in an external shock as the initially ultra-relativistic flow decelerated due to its interaction with the interstellar medium \citep{gia11}, such as in models that explain radio emission in gamma-ray burst afterglows \citep[e.g.][]{ber12, zau13}. The complex behaviour of the radio lightcurves is indicative of a forward shock driven by a jet with a fast core plus slow sheath structure \citep[][]{mim15}.   

If the non-thermal emission came from a highly anisotropic component, such as a relativistic jet formed in {\it{Swift}}~J1644$+$5734, the high luminosity up to $10^{48}$~erg\,s$^{-1}$ in X-rays could be naturally explained as a result of strong beaming effects \citep[e.g.][]{bur11}. The existence of a relativistic jet in {\it{Swift}}~J1644$+$5734 may be confirmed directly via measuring its apparent jet speed. The first very long baseline interferometry (VLBI) astrometry was performed by \citet{ber12} using the Very Long Baseline Array (VLBA) and Effelsberg 100~m radio telescope at 22 and 8.4~GHz. The high-frequency VLBI observations from the first half year gave a $3\sigma$ upper limit of 3.8$c$ on the apparent expansion velocity. At the same time, we initiated another VLBI campaign with the European VLBI Network (EVN) at 5~GHz. In our campaign, a relatively lower angular resolution at a longer wavelength is compensated by a longer time baseline. Moreover, an in-beam phase-referencing source is possible due to a wider antenna primary beam at the longer wavelength.

The paper is organised in the following sequence. We introduce our EVN observations and data calibration in Section~\ref{sec2}. We present our extremely high precision VLBI astrometry results in Section~\ref{sec3}, and a firm upper limit on the apparent jet speed in Section~\ref{sec4}. 

\section{Observations and Calibration}
\label{sec2}

\begin{table}
\caption{Summary of the EVN observations of {\it{Swift}}~J1644$+$5734. } \label{tab1}
\scriptsize
\setlength{\tabcolsep}{5pt}
\begin{tabular}{lclr}
\hline
Project & Date       & Participating EVN stations$^a$                                &  Duration     \\
Name    & yy-mm-dd   &                                                               & (hour)        \\
\hline
RP017   & 11-04-12   & EfWbJb2OnTrYsMc\phantom{NtSvZcBdUr}Sh                         &    9.0        \\
EY018A  & 12-03-08   & EfWbJb1OnTrYsMc\phantom{Nt}SvZcBdUrSh                         &   12.0        \\
EY018B  & 12-05-31   & EfWbJb1OnTrYsMcNtSvZcBdUr\phantom{Sh}                         &   12.0        \\
EY019   & 13-02-12   & \phantom{Ef}WbJb1OnTrYsMcNtSv\phantom{ZcBd}UrSh               &   11.5        \\
EY020A  & 13-10-21   & EfWbJb1OnTrYsMcNtSvZcBdUrSh                                   &   12.0        \\
EY020B  & 15-03-09   & EfWbJb1OnTrYsMcNtSvZcBd\phantom{Ur}Sh                         &   12.0        \\
\hline
\end{tabular} \\
$^a$ Ef: Effelsberg, Wb: Westerbork in the phased-array mode, Jb1: Jodrell Bank Lovell, Jb2: Jodrell Bank MKII, On: Onsala, Tr: Torun, Ys: Yebes, Mc: Medicina, Nt: Noto, Sv: Svetloe, Zc: Zelenchukskaya, Bd: Badary, Ur: Urumqi and Sh: Shanghai. 
\end{table}

Our EVN observations of {\it{Swift}}~J1644$+$5734 are summarised in Table~\ref{tab1}. All the VLBI experiments were observed at 5~GHz with a recording rate of 1024 Mbps (dual polarisation, $8\times16$ MHz per polarisation, 2-bit quantization). Note that the first experiment was done with the e-VLBI technique which enables real-time data transfer from stations to the central correlator via optical fibre cables. One of the goals in the first observing run, besides rapid VLBI detection of the transient, was to verify the compactness of a nearby ($2\farcm9$ away from the main target) radio source, FIRST J1644$+$5736, with a total flux density of $0.75\pm0.15$~mJy at 1.4 GHz \citep{bec95} and a flat spectrum based on the Westerbork Synthesis Radio Telescope (WSRT) image of the transient field \citep{lev11}. The faint source FIRST J1644$+$5736 is quite compact with a peak brightness of 0.7~mJy/beam at VLBI resolution. It is becoming a common practice for VLBI users to select an additional nearby phase-referencing source in e-VLBI runs to improve astrometric accuracy \citep[e.g.][]{par13}.

\begin{figure}
  \centering
  \includegraphics[width=0.47\textwidth]{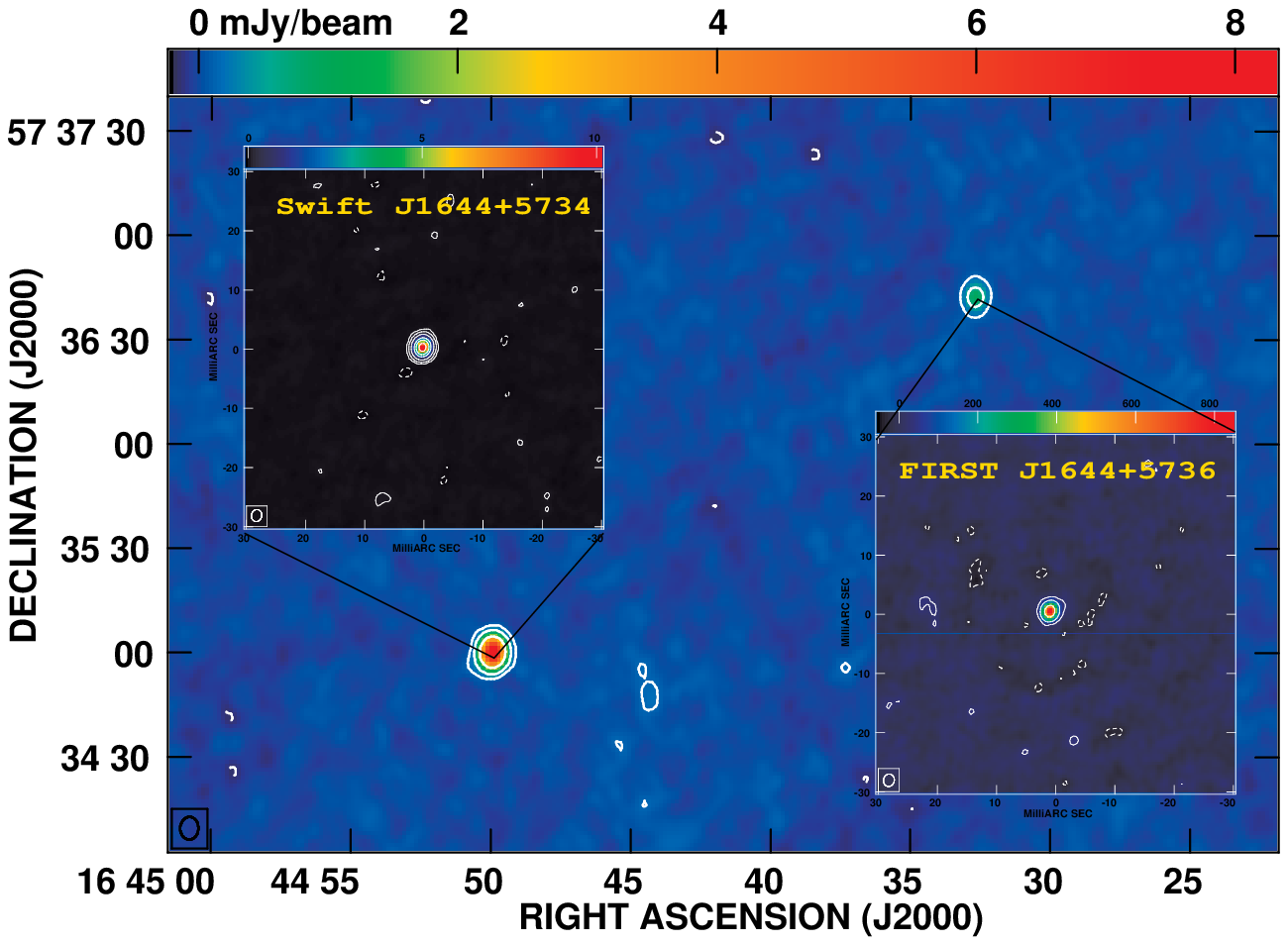} \\
  \includegraphics[width=0.45\textwidth]{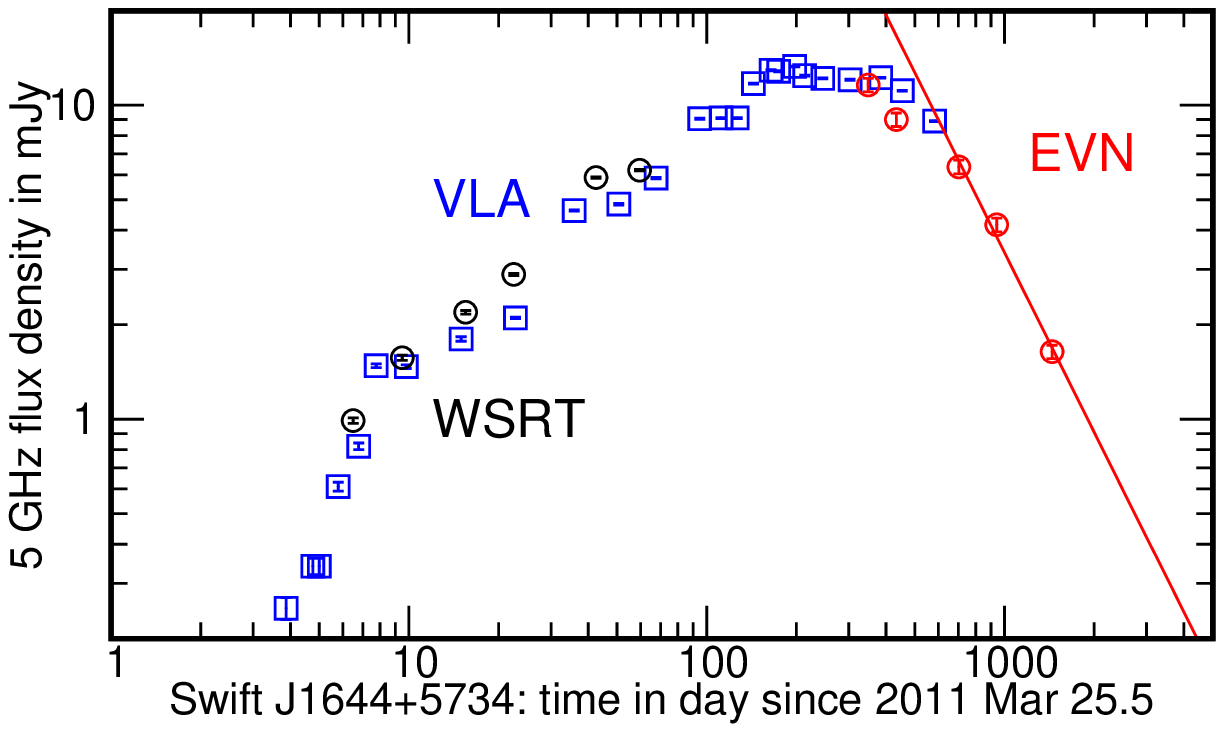} 
\caption{Top: The natural weighted EVN images of the tidal disruption event {\it{Swift}}~J1644$+$5734 and the phase-referencing source FIRST~J1644$+$5736 on 2012 Mar 8 in the background WSRT map. Bottom: The radio light curve at 5~GHz measured by the the EVN (red), the VLA \citep[blue,][]{ber12, zau13} and the WSRT \citep[black,][]{wie12}. The red line is the power law decay model ($S_\mathrm{5GHz}\propto t^{-1.9\pm0.2}_\mathrm{obs}$) that we fitted to the latest three points. }\label{fig1}
\end{figure}

The bright source ICRF2 J1638$+$5720 ($\sim$55$\arcmin$ away from the main target) was also observed to provide the traditional phase-referencing calibration and a reference position within the ICRF2 frame. The sub-mJy source FIRST J1644$+$5736 was only used as a nearby reference point using the inverse phase-referencing technique. We pointed small telescopes to the mid-point of the in-beam pair {\it{Swift}} J1644$+$5734 and FIRST J1644$+$5736 in order to observe both simultaneously. With the WSRT and Ef, we conducted nodding observations with a duty-cycle of $\sim$7 minutes ($\sim$1~min on the ICRF2 source, $\sim$2~mins on the TDE source, $\sim$4~mins on the FIRST source). In the first two epochs, Jb1 followed the strategy of Ef and WSRT while missing some scans because of its mechanical limit of $\leq$12 source changes per hour. In the subsequent three epochs, it followed the schedule of the smaller telescopes to achieve better phase calibration. The correlation was done with the EVN software correlator at JIVE \citep[SFXC;][]{kei15}. The correlation phase centres of the full EVN observations are listed in Table~\ref{tab2}.  

\begin{table}
\caption{The correlation phase centres in the EVN observations.} \label{tab2}
\scriptsize
\setlength{\tabcolsep}{7pt}
\centering
\begin{tabular}{rll}
\hline
Source  Name               &  $\alpha$ (J2000)                           & $\delta$ (J2000)                      \\  
\hline
{\it{Swift}}~J1644$+$5734  &   $16^\mathrm{h}44^\mathrm{m}49\fs9313$     &  $+57\degr34\arcmin59\farcs6893$      \\
FIRST~J1644$+$5736         &   $16^\mathrm{h}44^\mathrm{m}32\fs680951$   &  $+57\degr36\arcmin42\farcs33133$     \\
ICRF2 J1638$+$5720         &   $16^\mathrm{h}38^\mathrm{m}13\fs4562980$  &  $+57\degr20\arcmin23\farcs979046$    \\
\hline
\end{tabular}
\end{table}

The data were calibrated via a script developed by members of our team using Parseltongue \citep{ken06}, a Python interface to AIPS \citep[Astronomical Image Processing System;][]{gre03}. The a-priori amplitude calibration including the primary-beam correction was done in the same way as described by \citet{cao14}. Because of the imperfect antenna beam model, we expect to have a systematic flux density measurement error $\leq$15\,\% at the $3\sigma$ level. The ionospheric delay was corrected via total electron content measurements from GPS monitoring. Phase contributions from the antenna parallactic angle were removed before fringe-fitting. The fringe-fitting and the bandpass calibration were performed with ICRF2~J1638$+$5720. Finally, the data were averaged in each subband and split into single-source files. Note that the amplitude bandpass solutions were not applied to our data, to avoid introducing noise from the band-edge channels in the process of spectral averaging across the bands.

We imaged ICRF2~J1638$+$5720 manually in Difmap \citep{she94}. The calibrator clearly shows a linear jet structure with a total flux density around 1~Jy. We located the centroid of the jet base/radio core via Gaussian model fitting and then fixed it at the image origin. After that, we re-ran our script to remove the source structure. Both the amplitude and phase solutions were solved in the later self-calibration, and then applied to both the TDE and FIRST sources in AIPS. In the latter inverse phase-referencing calibration, we located the TDE source first by fitting a point source to the $uv$ data, and then iteratively ran self-calibration and point-source model fitting at the measured location in Difmap. As the TDE source was quite bright, amplitude self-calibration was also performed in the later steps with a long ($\geq$1 hour) solution interval. Using the final TDE source model reported in Table~\ref{tab3} and from the same epoch, we re-ran the self-calibration in AIPS to transfer its solutions to the FIRST source. With the additional self-calibration, the FIRST source shifted about 122~$\umu$as in the first e-EVN observations, and 18, 16, 7, 18, and 10~$\umu$as in the rest full EVN observations, as expected from their small phase solutions. No self-calibration was performed on the FIRST source. The final position offsets are summarised in Table~\ref{tab3}.

\section{High Precision VLBI Astrometry}
\label{sec3}

\begin{figure*}
  \centering
  \includegraphics[width=0.165\textwidth]{fig2a.eps}  
  \includegraphics[width=0.825\textwidth]{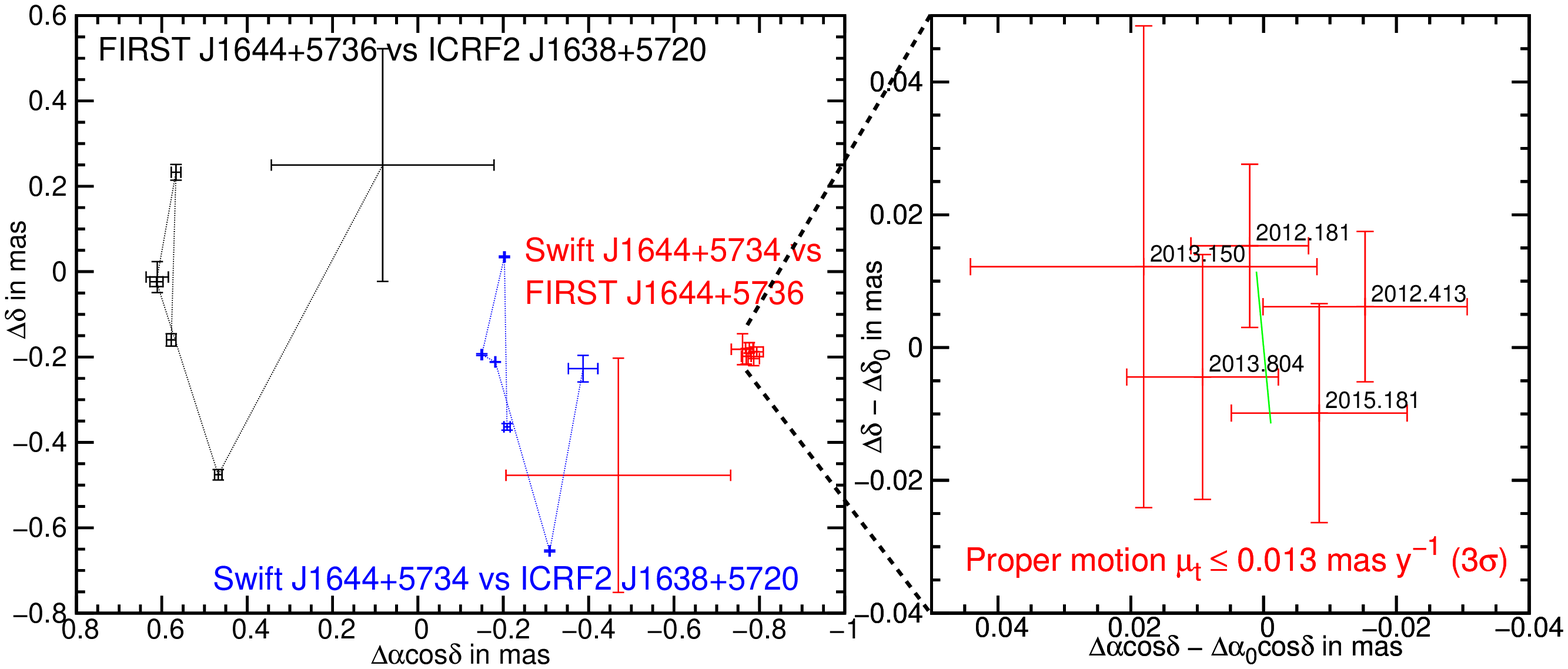}  \\
  \caption{Left: The EVN 5~GHz image of ICRF2~J1638$+$5720. Middle and Right: Plots of the relative VLBI position offsets of both {\it{Swift}}~J1644$+$5734 (blue) and FIRST~J1644$+$5736 (black) observed with respect to ICRF2~J1638$+$5720 (about 54$\arcmin$ away), and their differences (red), i.e. {\it{Swift}}~J1644$+$5734 with respect to FIRST~J1644$+$5736 (2$\farcm$9 away). The related position offsets and 1$\sigma$ error bars are listed in Table~\ref{tab3}. The green line is the proper motion model listed in Table~\ref{tab4}. }\label{fig2}
\end{figure*}

\begin{table*}
\caption{Summary of the EVN imaging results of {\it{Swift}}~J1644$+$5734 and FIRST~J1644$+$5736. The columns give (1) Modified Julian Day (MJD), (2) sizes of the beam major and minor axes, (3 -- 6) FIRST J1644$+$5736: total flux density ($S_\mathrm{5GHz}$), image signal to noise ratio (SNR), relative position offsets ($\Delta\alpha\cos\delta$, $\Delta\delta$), (7 -- 10) the same parameters for {\it{Swift}} J1644$+$5734, (11) size estimation. All the errors are 1$\sigma$. The offsets were measured by fitting a point source to the $uv$ data and with respect to the image origin listed in Table~\ref{tab2}.  } \label{tab3}
\scriptsize
\setlength{\tabcolsep}{4.0pt}
\centering
\begin{tabular}{cc  rrcc c rrccc}
\hline
            &       &  \multicolumn{4}{|c|}{FIRST~J1644$+$5736 versus ICRF2 J1638$+$5720}  &  &   \multicolumn{4}{|c|}{{\it{Swift}}~J1644$+$5734 versus ICRF2 J1638$+$5720}                  \\ 
\cline{3-6}        \cline{8-12}
MJD         & FWHM  &  $S_\mathrm{5GHz}$                     
                              & SNR &   $\Delta\alpha\cos\delta$  
                                                        & $\Delta\delta$     &  & $S_\mathrm{5GHz} $ & SNR &  $\Delta\alpha\cos\delta$   
                                                                                                                             &  $\Delta\delta$   & $\theta_\mathrm{size}$ \\ 
(day)       & (mas)           &  (mJy)         &     & (mas)             & (mas)              &  &   (mJy)            &      &  (mas)             & (mas)             & (mas)    \\
\hline 
55664.048   & $7.1\times3.0$  &  $0.64\pm0.03$ &   8 & $+0.083\pm0.261$  &  $+0.250\pm0.273$  &  &    $3.96\pm0.20$   &  72 &  $-0.387\pm0.035$  &  $-0.227\pm0.031$ & $0.35\pm0.17$  \\ 
55994.255   & $2.2\times1.8$  &  $0.83\pm0.04$ &  76 & $+0.468\pm0.009$  &  $-0.476\pm0.012$  &  &   $11.60\pm0.58$   & 537 &  $-0.309\pm0.001$  &  $-0.654\pm0.002$ & $0.12\pm0.06$  \\
56079.026   & $2.2\times1.9$  &  $0.69\pm0.03$ &  85 & $+0.612\pm0.015$  &  $-0.024\pm0.011$  &  &    $9.00\pm0.45$   & 241 &  $-0.181\pm0.001$  &  $-0.211\pm0.001$ & $0.15\pm0.07$  \\
56348.307   & $2.4\times1.8$  &  $0.67\pm0.03$ &  44 & $+0.611\pm0.026$  &  $-0.013\pm0.036$  &  &    $6.36\pm0.32$   & 285 &  $-0.150\pm0.002$  &  $-0.194\pm0.002$ & $0.20\pm0.08$  \\
56586.526   & $2.1\times1.6$  &  $0.78\pm0.04$ &  72 & $+0.567\pm0.011$  &  $+0.233\pm0.018$  &  &    $4.16\pm0.21$   & 244 &  $-0.203\pm0.002$  &  $+0.035\pm0.002$ & $0.14\pm0.06$  \\ 
57090.255   & $2.2\times1.9$  &  $0.66\pm0.03$ &  64 & $+0.578\pm0.011$  &  $-0.160\pm0.015$  &  &    $1.64\pm0.08$   & 146 &  $-0.209\pm0.007$  &  $-0.364\pm0.008$ & $0.24\pm0.11$  \\
\hline
\end{tabular}
\end{table*}

The EVN images of {\it{Swift}}~J1644$+$5734 and FIRST J1644$+$5736 observed on 2012 Mar 8 are shown in Fig.~\ref{fig1}. We had a nearly circular beam with a full width at half maximum (FWHM) of $\sim$2~mas using natural weighting. With an image sensitivity of $0.011$~mJy\,beam$^{-1}$ (1$\sigma$), we did not detect any secondary components or remnants from possible earlier jet activities. {\it{Swift}}~J1644$+$5734 was decaying quite rapidly over the last three epochs with $S_\mathrm{5GHz}\propto t^{-1.9\pm0.2}_\mathrm{obs}$. The fast decay was also observed at 15 GHz \citep{zau13}. The in-beam reference source FIRST J1644$+$5736 had an average flux density of 0.71~mJy and a standard deviation of~0.08 mJy. Comparing with our total flux density measurement error, $\leq$15\% (see Sec.~\ref{sec2}), there was no significant flux density variation observed in FIRST J1644$+$5736. 

According to our data reduction strategy, the relative position offsets in Table~\ref{tab3} were measured directly with the reference source ICRF2~J1638$+$5720 in case of {\it{Swift}}~J1644$+$5734, and indirectly via {\it{Swift}}~J1644$+$5734 as a bridge in case of FIRST J1644$+$5736. The 1$\sigma$ position error was derived from the statistical distribution of the fitting results of eight separate subband data. The position offsets and errors are shown in Fig.~\ref{fig2}. Both sources had the largest error bars in the first epoch because the first experiment mainly aimed to achieve a rapid detection of them using the e-VLBI technique. If we exclude the two points with the largest error, both sources have a distribution quite resembling each other and a scatter (standard deviation: 59~$\umu$as in R.A and 260~$\umu$as in Dec.) much larger than their respective formal error bars. Comparing with the jet direction in ICRF2~J1638$+$5720, we notice that the largest scattering direction is roughly aligned with the jet direction. Therefore, the jet base in ICRF2~J1638$+$5720 appears to vary due to its intrinsic structure changes. Jitter in the radio core position of ICRF2 sources is frequently seen \citep{moo11}, and this significantly contributes to the astrometry uncertainties when ICRF2 are used as calibrators in phase-referencing VLBI observations (at multiple epochs). There was also an uncorrected systematic phase error caused by the propagation effects in the ionosphere and the troposphere \citep{rei14}. The systematic position error in R.A. is in agreement with the error analysis presented by \citet{pra06} via simulating VLBI data and considering various error components.  
     
The red points in Fig.~\ref{fig2} represent the relative position measurements with respect to FIRST~J1644$+$5736 ($2\farcm9$ apart). The first low-SNR point is excluded in our analysis. All the remaining high-SNR data points are well located in a rather compact region. The zoom-in plot of the region is shown in the right panel. The standard deviation of the five data points is 13~$\umu$as in R.A. and 11~$\umu$as in Dec., about 1/160 of the synthesised beam. Since the scattering is as small as their formal error bars, the systematic position error has been reduced significantly to a low level ($<$10~$\umu$as). As far as we know, this position precision is one of the best phase-referencing measurements \citep[see a review by][]{rei14}. Currently, there are only a few astrometry projects carried out with the EVN, such as high-precision pulsar parallax measurements by \citet{kir15} at 5 GHz and \citet{du14} at 1.6 GHz. These recent VLBI studies, and ours, show that the EVN astrometry can achieve a precision at the top level. 

To search for possible proper motion, we fit the five data points to a linear function and list the best-fit parameters and their asymptotic standard error in Table~\ref{tab4}. Our fitting gives a total proper motion $\mu_\mathrm{t}=7.6\pm4.2$~$\umu$as\,yr$^{-1}$ along position angle $\theta_\mu=-175\degr\pm26\degr$. If our error bars are scaled by a small factor 0.68, then the reduced $\chi^2_\mathrm{red}=1$. Since $\mu_\mathrm{t}$ is estimated at $1.8\sigma$ significance, no proper motion can be declared as detected here. Our proper motion measurement was not affected by the opacity variation in the TDE source, because the synchrotron-self absorption frequency was below 5~GHz \citep{zau13}. Note that the proper motion measurement also includes a contribution from the FIRST source. However, its fraction is likely small given the lack of detectable flux variability. 
     
The source {\it{Swift}}~J1644$+$5734 was highly compact during our observations. To give a quantitative description of its compactness, we also fitted the visibility data to a circular Gaussian model in Difmap. The last column in Table~\ref{tab3} gives these model-dependent estimations and Monte Carlo 1$\sigma$ error. In our simulations, we repeatedly added a point source with the same flux density as that of {\it{Swift}}~J1644$+$5734 at a random position into the calibrated visibility data, in which the real source {\it{Swift}}~J1644$+$5734 was removed, and then fitted them to a circular Gaussian model in Difmap. Comparisons with our Monte Carlo 3$\sigma$ errors (purely VLBI configuration-dependent limits) show that all our size measurements are not significant ($<3\sigma$). Thus, the source was not resolved in any epoch.

\section{No apparent superluminal motion}
\label{sec4}
\begin{figure}
  \centering
  \includegraphics[width=0.42\textwidth]{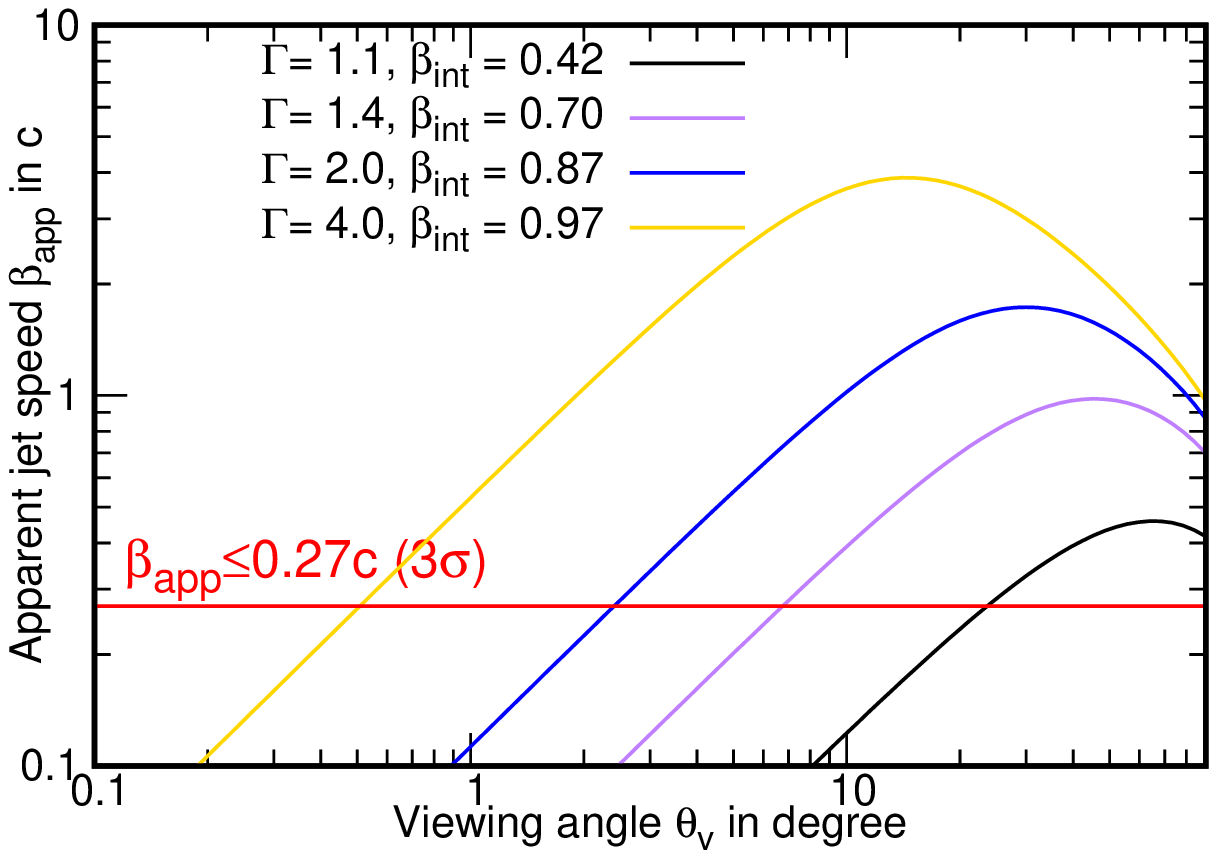}
\caption{The VLBI constraints on the viewing angle and intrinsic jet speed. The blue line represents the intrinsic jet speed inferred by \citet{zau13} during our observations. }\label{fig3}
\end{figure}

The apparent jet speed $\beta_\mathrm{app}$ depends on the intrinsic jet speed $\beta_\mathrm{int}$ and the viewing angle $\theta_\mathrm{v}$ via the known equation $\beta_\mathrm{app}=\beta_\mathrm{int}\sin\theta_\mathrm{v}(1-\beta_\mathrm{int}\cos\theta_\mathrm{v})^{-1}$. The dependence is plotted in Fig.~\ref{fig3} for different values of $\theta_\mathrm{v}$, $\beta_\mathrm{app}$ and $\beta_\mathrm{int}$, with the bulk Lorentz factor $\Gamma=(1-\beta^2_\mathrm{int})^{-1/2}$. In case of a VLBI detection of a large superluminal motion, one can provide a lower limit on $\Gamma$, $\Gamma_\mathrm{min}=(\beta_\mathrm{app}^2+1)^{1/2}$, and an upper limit on $\theta_\mathrm{v}$, $\cos\theta_\mathrm{v,max}=(\beta_\mathrm{app}^2-1)(\beta_\mathrm{app}^2+1)^{-1}$ \citep[e.g. ][]{bot12}.  

\begin{table}
\caption{The best-fit parameters in the proper motion model.} \label{tab4}
\scriptsize
\setlength{\tabcolsep}{5pt}
\centering
\begin{tabular}{ll}
\hline
Fitting parameters          &   Best-fit value                                  \\  
\hline
Total proper motion         &   $\mu_\mathrm{t}=0.0076\pm0.0042$~mas\,y$^{-1}$  \\
Position angle              &   $\theta_\mathrm{\mu}=-175\pm26\deg$             \\
Offset in R.A. at 2013.68   &   $\Delta\alpha_0\cos\delta=-0.778\pm0.004$~mas   \\
Offset in Dec. at 2013.68   &   $\Delta\delta_0=-0.194\pm0.005$~mas             \\
Reduced $\chi^2$            &   $\chi^2_\mathrm{red}=0.47$                      \\        
\hline
\end{tabular}
\end{table}   

Interestingly, there was no apparent superluminal motion observed in {\it{Swift}}~J1644$+$5734 during our EVN observations. According to our VLBI proper motion measurements and the redshift of {\it{Swift}}~J1644$+$5734, the apparent jet speed corrected for time dilation was $\beta_\mathrm{app}=(1+z)d_\mathrm{a}\mu_\mathrm{t}=0.16\pm0.09c$, where $c$ is the light speed and the angular-diameter distance $d_\mathrm{a}=1006$~Mpc at $z=0.354$ with the cosmological parameters $H_0=70$~km\,s$^{-1}$Mpc$^{-1}$, $\Omega_\mathrm{m}=0.28$, $\Omega_\mathrm{\Lambda}=0.72$. The 3$\sigma$ upper limit of $\beta_\mathrm{app}$ was 0.27$c$ between 2012 Mar 8 and 2015 Mar 9. The new EVN limit is significantly tighter than the early VLBA+Ef limit \citep[$\sim4c$ at 3$\sigma$ level,][]{ber12}. This is as expected since the EVN observations had a six times longer span and the reference source was quite stable.    

By analogy to GRB jets, \citet{zau13} modelled the broad band radio spectra and reported $\Gamma\simeq2$ at $t\sim582$ day (between our 2nd and 3rd full EVN observations), corresponding to the blue curve in Fig.~\ref{fig3}. Together with our VLBI constraint on $\beta_\mathrm{app}$ (red line), the radio jet, associated with the expanding forward shock, will be very close to the line of sight with a viewing angle $\theta_\mathrm{v}<3\degr$. In case of a small $\theta_\mathrm{v}$, the observed low $\beta_\mathrm{app}$ may also partly result from the average of $\beta_\mathrm{app}$ across a jet opening angle of just a few degrees \citep{gop04}. Alternatively, the radio jet might have a $\Gamma$ lower than the estimation by \citet{zau13}, i.e. $\Gamma<2$, due to a denser surrounding medium. This will give a much larger viewing angle, more favourable within the frame of the spine and sheath jet structure \citep{mim15}. In either case, the jet has decelerated to mildly relativistic speed within $t_\mathrm{obs}/(1+z)\sim3$~years after the trigger, implying that its Sedov length $R_\mathrm{s}=(3E_\mathrm{iso})^{1/3}(4\pi n_\mathrm{cnm}m_\mathrm{p}c^2)^{-1/3}\leq1$~pc, where the isotropic equivalent energy of the blast $E_\mathrm{iso}\sim10^{54}$~erg \citep{zau13} and $n_\mathrm{cnm}$ is the density of the circum-nuclear medium at distance $\sim$1~pc from the black hole. This constrains $n_\mathrm{cnm}\geq 5E_\mathrm{iso,54}$~cm$^{-3}$. The high $n_\mathrm{cnm}$ constraint also supports the explanation of the later X-ray light curve as a result of a Compton echo, as suggested by \citet{che16}.         

It is also possible that the radio jet in {\it{Swift}}~J1644$+$5734 was mildly relativistic, e.g. $\leq$\,0.3$c$ revealed by the simulation of a purely radiatively driven jet \citep{sad15}. However, it remains to be shown that such a model can account for the broad-band radio lightcurves.

Ultra high precision VLBI astrometry of new-borne jets, as demonstrated here with respect to a nearby faint source, will provide us with the most direct insights on the early phase of their formation and evolution. With more sensitive radio telescopes \citep[][]{par15}, e.g. Sardinia, Tianma, the upcoming FAST (Five hundred meter Aperture Spherical Telescope) and SKA1-MID, VLBI astrometry precision will reach $\umu$as level and enable us to shed new light on questions triggered by {\it{Swift}}~J1644$+$5734.

\section*{Acknowledgments}
\label{ack}
{\footnotesize We thank Thomas Krichbarum for critically reading the manuscript. T. An thanks the grant support from the 973 programme (No. 2013CB837900). The European VLBI Network is a joint facility of independent European, African, Asian, and North American radio astronomy institutes. Scientific results presented in this publication are derived from the following EVN projects: RP017, EY018, EY019, and EY020. The study has received funding from the European Commission Seventh Framework Programme (FP/2007-2013) under grant agreement No. 283393 (RadioNet3).  e-VLBI research infrastructure in Europe is supported by the European Union's Seventh Framework Programme (FP7/2007-2013) under grant agreement number RI-261525 NEXPReS. The WSRT is operated by ASTRON (Netherlands Institute for Radio Astronomy) with support from the Netherlands Organisation for Scientific Research. This research has made use of the NASA/IPAC Extragalactic Database (NED) which is operated by the Jet Propulsion Laboratory, California Institute of Technology, under contract with the National Aeronautics and Space Administration.}

\vspace{-12pt}

\bsp  
\label{lastpage}
\end{document}